# A Taylor Series Approximation Model for Characterizing the Output Resistance of a GFET

Xiomara Ribero-Figueroa, Anibal Pacheco-Sanchez, Tzu-Jung Huang, David Jiménez, Ivan Puchades, and Reydezel Torres-Torres, *Senior Member, IEEE*

*Abstract*—The mobility-degradation-based model for the drain-to-source or output resistance of a graphene field-effect-transistor is linearized here using a Taylor series approximation. This simplification is shown to be valid from magnitudes of the gate voltage not significantly higher than the Dirac voltage, and it enables the analytical determination of the transconductance parameter, the voltage related to residual charges, and a bias-independent series resistance of the GFET. Furthermore, a continuous representation of the device's static response is achieved when substituting the extracted parameters into the model, regardless the transfer characteristic symmetry with respect to the Dirac voltage.

## I. INTRODUCTION

Graphene field-effect transistors (GFETs) are gaining importance for various applications [1]–[3], including microwave circuits [4]. This is mainly due to the high carrier mobility and high driven current capabilities involving a low resistance [5]. Thus, analyzing their behavior is crucial for circuit-oriented modeling, and device optimization. In this regard, a primary tool for characterization is direct current (DC) equipment. Therefore, DC methods are typically employed to characterize the electrical transport performance of these devices. Nonetheless, traditional approaches to characterize the GFET resistances involve regressions on data measured to transistors of different lengths [6]–[7], which might include errors due to the variability of device characteristics even when they exhibit identical layout and are fabricated within the same die. Hence, a methodology for extracting GFET model parameters from single device measurements is desirable.

Manuscript received March 21, 2023. This work was supported in part by Consejo Nacional de Ciencia y Tecnología (CONACyT), Mexico through Scholarship 288875, from European Union's Horizon 2020 research and innovation programme under grant agreement No GrapheneCore3 881603, from Ministerio de Ciencia, Innovación y Universidades under grant agreements PID2021-127840NB-I00 (MCIN/AEI/FEDER, UE). *Corresponding author: X. Ribero-Figueroa*

Xiomara Ribero-Figueroa and Reydezel Torres-Torres are with INAOE, Puebla 72840, Mexico (e-mail: xiomis.93@gmail.com).

A. Pacheco-Sanchez and David Jiménez are with the Departament d'Enginyeria Electrònica, Escola d'Enginyeria, Universitat Autònoma de Barcelona, Bellaterra 08193, Spain.

Tzu-Jung Huang, Department of Microsystems Engineering, Rochester Institute of Technology, Rochester, NY 14623, USA.

Ivan Puchades, Department of Electrical and Microelectronic Engineering, Rochester Institute of Technology, Rochester, NY 14623, USA.

Color versions of one or more of the figures in this paper are available online at http://ieeexplore.ieee.org.

Digital Object Identifier xx

A further complication in modeling and characterization, arises from ambipolar response of GFETs, which includes both hole and electron conduction [8]. Thus, when implementing the model for its output resistance requires ensuring a smooth transition between these two regimes. To resolve this issue, several methods have been reported [6], [9]–[12], whereas iterative techniques are available to achieve a quasi-continuous response [9]. Alternatively, transcendent functions have been combined with the model to smooth the representation of the output resistance at the transition between the hole and electron dominated regions [10]. Nonetheless, this increase in the difficulty of the model may be unnecessary when analytically determining the model parameters.

Here, the linearization of the square-root term involved in the model for the GFET's output resistance is proposed to dramatically simplify the extraction of the corresponding parameters from DC curves conducted on a single device. This avoids using iterative methods or methods that require multiple devices. In fact, the achieved individual extraction of parameters is useful, for instance, to quantify the contributions related to electron and hole transport to the bias independent resistance. This is relevant not only when implementing circuit-oriented models but also during technology development, for example, to provide feedback on performance, adjust process variables, and assess yield.

## II. OUTPUT RESISTANCE MODEL PARAMETER EXTRACTION

The output resistance, $R_{DS}$, of the GFET is composed of the channel resistance ($R_{CH}$) and parasitics associated with the extrinsic source and drain resistances. Effectively, however, it is convenient to express $R_{DS}$ as the sum of two components: one representing the part of the graphene channel controlled by the gate bias ($R_{BIAS}$) and the other independent of it ($R_{CON}$). Here, the expression from [13] applies:

$$R_{DS} = R_{CON} + R_{BIAS} = R_{CON} + \frac{1}{\beta\sqrt{V_{GS0}^2 + V_0^2}} \quad (1)$$

where $\beta = (W/L)\mu_0 C'_{ox}$ represents the transconductance parameter, $C'_{ox}$ is the per-unit-area gate oxide capacitance, $\mu_0$ is the low-field mobility, and $W$ and $L$ are the channel width and length, respectively. $V_{GS0}$ is defined in terms of the Dirac voltage ($V_{Dirac} = V_{GS}$ at maximum $R_{DS}$) as $V_{GS0} = V_{GS}-V_{Dirac}$, whereas $V_0$ is associated with a residual charge density and is considered in the model as a fitting parameter. Also,

$$R_{CON} = R_C + \frac{\theta_{ch}}{\beta} \quad (2)$$

where $\theta_{ch}$ is the mobility degradation coefficient, whereas $R_C$ is the resistance associated with the drain and source terminals.

The second term on the right in (1) can be rewritten as:

$$\frac{1}{\beta}(V_{GS0}^2 + V_0^2)^{-\frac{1}{2}} = \frac{1}{\beta V_{GS0}}\left(1 + \frac{V_0^2}{V_{GS0}^2}\right)^{-\frac{1}{2}}$$
$$\cong \frac{1}{\beta V_{GS0}}\left(1 - \frac{V_0^2}{2V_{GS0}^2}\right) \quad (3)$$

where the square root was linearized using the first two terms of a Taylor series expansion, considering $V_0^2 \ll V_{GS0}^2$. This assumption holds true at gate biases where channel conduction is dominated by one type of charge carrier, which occurs when $|V_{GS0}| \gg 0$. Substituting (3) into (1) in this scenario yields:

$$R_{DS} \approx \left[R_{CON} + \frac{1}{\beta |V_{GS0}|}\left(1 - \frac{V_0^2}{2V_{GS0}^2}\right)\right]_{|V_{GS0}| \gg 0} \quad (4)$$

In (4), $V_{GS0}$ can be straightforwardly obtained at any $V_{GS}$ once $V_{Dirac}$ is determined by analyzing the voltage at which $R_{DS}$ reaches its maximum magnitude. This leaves three unknowns in a single equation, which can be reduced to two by applying a derivative function. For this purpose, the following auxiliar parameter can be defined:

$$y = \frac{1}{V_{GS0}}\frac{\partial(V_{GS0}^3 R_{DS})}{\partial V_{GS0}} = 3R_{CON}V_{GS0} + \frac{2}{\beta} \quad (5)$$

Based on this equation, $y$ can be calculated from experimental data and plotted against $V_{GS0}$. Thus, considering $m$ as the slope and $b$ as the y-intercept of the linear regression of these data, $R_{CON} = m/3$ and $\beta = 2/b$ are calculated. As mentioned, the $V_{GS0} \gg 0$ V condition should be maintained when generating this plot.

The remaining unknown is $V_0$, which is obtained at $V_{GS} = V_{Dirac}$ (i.e., at $V_{GS0} = 0$), resulting in $R_{DS} = R_{Dirac}$. Under this assumption, it is possible to solve (1) for $V_0$ to yield:

$$V_0 = \frac{1}{\beta(R_{Dirac} - R_{CON})} \quad (6)$$

The key advantage here is the requirement of only one derivative, minimizing noise in experimental data compared to methods that rely on multiple derivatives or iterative calculations. Consequently, these methods are reserved for scenarios where model complexity increases due to high-order effects and involves additional parameters. In fact, the proposal is also applicable to some of these cases, for example, when implementing circuit-design representations for large-signal operation, built on a solid DC modeling foundation.

### III. EXPERIMENTAL VERIFICATION

The response of a GFET is considered symmetrical when the $R_{DS}$ versus $V_{GS0}$ curve exhibits the characteristics of an even function. In this case, it can be assumed that the hole-dominated current for $V_{GS0} < 0$ equals the electron-dominated current for $V_{GS0} > 0$. Nevertheless, this condition is difficult to achieve in practice mainly due to the different mobility exhibited by the two types of charge carriers within the graphene channel [14]–[15]. This section explains the application of the proposed methodology to GFETs that exhibit a quasi-symmetrical response due to reduced device degradation when measured under vacuum conditions [16], and subsequently to GFETs measured under ambient conditions where the response shows significant asymmetry. This illustrates the usefulness of the proposal not only when similar responses are obtained for the device under hole-dominated and electron-dominated conduction regions but also when noticeable asymmetry is observed, which is commonly encountered in practice.

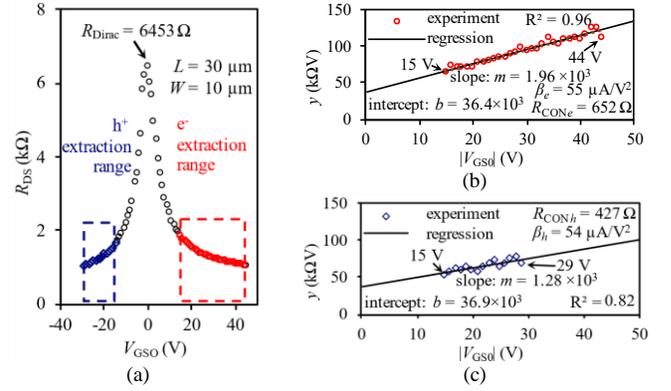

Fig. 1. Experimental data corresponding to a GFET with quasi-symmetrical response. a) $R_{DS}$ versus $V_{GS0}$ curve [16], and linear regressions to extract $\beta$ and $R_{CON}$ for b) electron-dominated, and c) hole-dominated conduction regions.

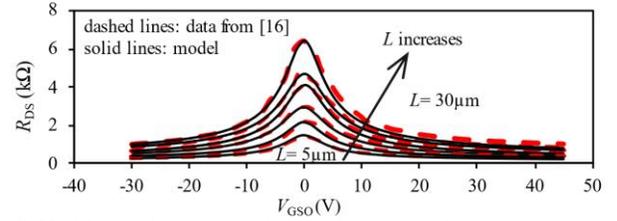

Fig. 2. Model given by equation (1) after determining the parameters using the proposed method, compared to experimental data. The data correspond to GFETs varying in length from 5 μm to 30 μm in increments of 5 μm.

### A. Devices with quasi-symmetric response

Fig. 1(a) shows the quasi-symmetrical response of a GFET with channel width $W = 10$ μm and length $L = 30$ μm, as reported in [16]. The measurements were conducted under vacuum conditions, as specified, to minimize the influence of factors that can uncontrollably alter the carrier mobility in graphene, such as humidity and other residues.

To start the parameter extraction, $R_{CONe}$ and $\beta_e$ are initially obtained, where '$e$', added to the original subscript, denotes that these parameters correspond to the electron-dominated region. Hence, based on (5), experimental data within the voltage range $15\,\text{V} \leq V_{GS0} \leq 44\,\text{V}$ are plotted as shown in Fig. 1(b). Excellent linearity is observed, which allows determining $R_{CONe}$ and $\beta_e$. In a similar fashion, $R_{CONh}$ and $\beta_h$, with '$h$' added to the subscript to refer to parameters for the hole-dominated region, are extracted as illustrated in Fig. 1(c). Afterwards, by considering $R_{Dirac} = 6453$ Ω obtained from the peak of the curve in Fig. 1(a), $V_{0e}$ is calculated applying (6), using the determined data for $R_{CONe}$ and $\beta_e$. Likewise, $V_{0h}$ is obtained using $R_{CONh}$ and $\beta_h$.

Once the model parameters were obtained with the aid of the approximation in (4), the original model in (1) is employed to obtain the electron-dominated $R_{DS}$ curve for $V_{GS0} \geq 0$ V considering the values for $R_{CONe}, \beta_e$, and $V_{0e}$. For $V_{GS0} \leq 0$ V the parameters for the hole-dominated region are used. The results for GFETs varying in length from 5 μm to 30 μm in increments of 5 μm are illustrated in Fig. 2. Notice the model accuracy, which validates the proposed extraction methodology for devices with quasi-symmetrical response. Furthermore, no additional functions to transition between hole-and electron-dominated regions (e.g., [10]) were required to maintain the model continuous within the full voltage range.

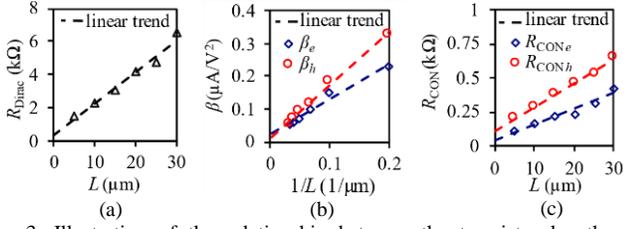

Fig. 3. Illustration of the relationship between the transistor length and extracted parameters a) $R_{Dirac}$, b) $\beta_h$ and $\beta_e$ and c) $R_{CONh}$ and $R_{CONe}$.

TABLE I
$V_0$ FOR THE HOLE AND ELECTRON-DOMINATED REGIONS VARYING THE DEVICE LENGTH

| $L$ (μm) | 5 | 10 | 15 | 20 | 25 | 30 | mean. | std. dev. |
|---|---|---|---|---|---|---|---|---|
| $V_{0h}$ (V) | 3.16 | 3.24 | 3.7 | 3.44 | 3.62 | 3.06 | 3.37 | 0.26 |
| $V_{0e}$ (V) | 2.37 | 2.81 | 3.24 | 2.82 | 3.39 | 3.14 | 2.96 | 0.37 |

Verifying the scalability of the determined model parameters is also necessary. For instance, Fig. 3(a) and 3(b) show that $R_{Dirac}$ linearly increases with $L$, while $\beta_e$ and $\beta_h$ exhibit an inversely proportional relationship with it, as expected. On the other hand, the linear trend observed for the $R_{CONe}$ and $R_{CONh}$ versus $L$ data in Fig. 3(c) is due the inverse dependence of the channel resistance on the transconductance, as expressed in (2). Finally, the extracted $V_{0e}$ and $V_{0h}$ for the different transistor lengths are listed in Table I, showing their statistical mean and standard deviation. Since $V_0$ is involved with a residual charge concentration, it is different for each transistor and type of charge carrier. This points out one of the benefits of the proposed methodology, which is avoiding the assumption of parameter invariability for using arrays of devices to perform the corresponding extraction [17].

### B. Devices with asymmetric response

To further verify the usefulness of the methodology, GFETs were fabricated for measurement under ambient conditions, thereby favoring an asymmetrical device response. In this regard, lithography and lift-off steps were performed on a silicon wafer. Hence, the first step in this process was growing 90 nm of $SiO_2$ at 1000°C. Afterwards, 200 nm of aluminum was thermally evaporated and patterned through a lift-off lithography process to serve as a gate electrode. Then, a 45 nm layer of tetraethyl orthosilicate (TEOS) was deposited through plasma-enhanced chemical vapor deposition (PECVD) directly on the gate electrodes to serve as an alumina etch stop when opening contact vias. Subsequently, 15 nm of $Al_2O_3$ was deposited using an atomic layer deposition (ALD) system. At this point, a graphene monolayer sheet was transferred onto the wafer surface, which was then patterned through lithography. After this step, Ni and Au (5 nm and 95 nm, respectively) were thermally evaporated to form the source and drain contacts. Finally, the wafers were coated with photoresist, and contact vias were patterned and etched to expose the aluminum gate contact pads. All fabricated devices present $W$ = 11 μm and $L$ = 10 μm. Refer to [18] for a more detailed explanation of the manufacturing process of the transistors analyzed here.

The devices are integrated between two ground-signal-ground (GSG) pad arrays and have two fingers. Therefore, for the purpose of conducting DC measurements, four probes were used: one for the gate, one for the drain, and two for the source pads, as illustrated in Fig. 4. Measurements were taken using a semiconductor device analyzer (SDA) sweeping $V_{GS}$ from −8 V to 8 V, while maintaining a constant $V_{DS}$ = 1 V.

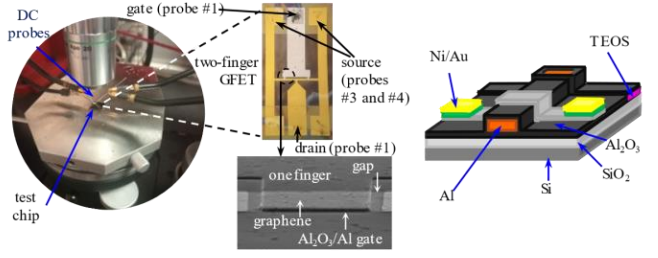

Fig 4. Experimental setup used to perform DC measurements, and device-under-test for $W$ = 11 μm and $L$ = 10 μm.

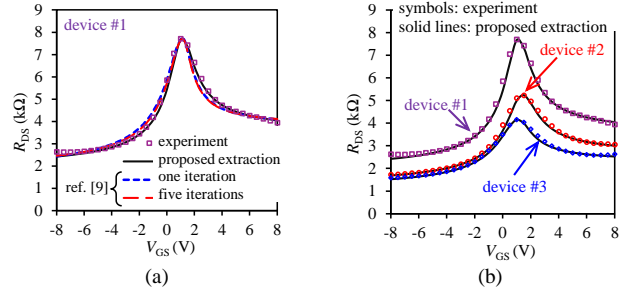

Fig. 5. Comparison of the model given by (1), implemented using the proposed methodology, with experimental data at $V_{DS}$ = 1 V: a) for one of the devices, comparing it with the method from [9], and b) using the proposed method for all three devices considered.

Fig. 5(a) shows the agreement with experimental data for $R_{DS}$ in one of the fabricated asymmetric devices using the proposed extraction methodology. Also, note in this figure that while the iterative method reported in [9] achieves acceptable accuracy, it does not match the level of accuracy of the proposed method and shows no improvement after 5 iterations for the considered device. Figure 5(b) shows the results for all fabricated devices using the proposed methodology. Notice the continuity attained at the Dirac point, effectively capturing the transition between regions dominated by holes and those dominated by electrons. Interestingly, although these devices have identical layout and were fabricated within the same test die, differences in their responses and extracted model parameters can be observed in Fig. 6. This variation is not uncommon in GFETs, where defects induced during the manufacturing process can affect hole and electron mobilities differently among devices within a single die. Hence, one of the applications of the proposed approach is the individual characterization of multiple devices for inspecting variability within a test chip or wafer.

### IV. VERIFICATION OF THE PROPOSED APPROXIMATION

The linearization of the model expressed in (1) using Taylor series allowed to use (4) to greatly simplify the parameter extraction for the $R_{DS}$ model. Now, after performing the model implementation, these two models are confronted in Figures 7(a) and 7(b) to illustrate the agreement between the proposed approximation and the mobility-degradation-based model. Note that even for $V_{GS0}$ as small as a few volts, the approximation is excellent for both, hole and electron-dominated conduction regions. Thus, (4) can be used under this condition for extracting the model parameters. Then, substituting these parameters into (1) allows for obtaining $R_{DS}$ at any normal operation $V_{GS}$ range. Moreover, for first-order simulations and calculations, (4) can be confidently used, assuming the $V_{GS0} \gg 0$ V condition. To meet this criterion, the approximation should be applied at sufficiently negative $V_{GS}$ for hole conduction and sufficiently positive $V_{GS}$ for electron conduction, ensuring a monotonically decreasing trend in the curve. However, near the Dirac voltage, equation (1) should be used, while the Taylor series approximation should be reserved for the parameter extraction.

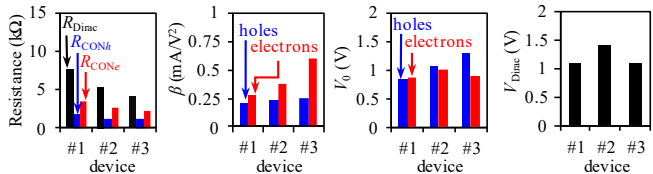

Fig. 6. Extracted model parameters for the GFETs with asymmetric response.

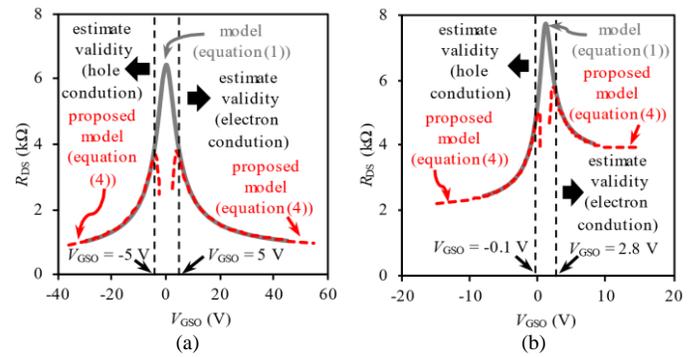

Fig. 7. Taylor series approximation model compared with the mobility-degradation-based model for the analyzed cases: a) symmetrical, and b) asymmetrical.

## V. CONCLUSION

The proposed approximated representation for the drain-to-source resistance of a GFET was used with success to extract the corresponding model parameters using data measured on a single device. It has been verified that the approximation is valid and significantly simplifies device characterization, when the experimental data are available at gate voltages somewhat higher than the Dirac voltage. This expression has been shown to be valid for GFETs with both symmetrical and asymmetrical response, allows to implement the mobility degradation-based model, and to continuously represent the device characteristics from hole-dominate to electron-dominated conduction regions.